# Data-Driven Modeling with Experimental Augmentation for the Modulation Strategy of the Dual-Active-Bridge Converter

Xinze Li, *Student Member, IEEE*, Josep Pou, *Fellow, IEEE*, Jiaxin Dong, *Student Member, IEEE*, Fanfan Lin, *Student Member, IEEE*, Changyun Wen, *Fellow, IEEE*, Suvajit Mukherjee, *Senior Member, IEEE*, and Xin Zhang, *Senior Member, IEEE*

***Abstract*—For the performance modeling of power converters, the mainstream approaches are essentially knowledge-based, suffering from heavy manpower burden and low modeling accuracy. Recent emerging data-driven techniques greatly relieve human reliance by automatic modeling from simulation data. However, model discrepancy may occur due to unmodeled parasitics, deficient thermal and magnetic models, unpredictable ambient conditions, etc. These inaccurate data-driven models based on pure simulation cannot represent the practical performance in physical world, hindering their applications in power converter modeling. To alleviate model discrepancy and improve accuracy in practice, this paper proposes a novel data-driven modeling with experimental augmentation (D²EA), leveraging both simulation data and experimental data. In D²EA, simulation data aims to establish basic functional landscape, and experimental data focuses on matching actual performance in real world. The D²EA approach is instantiated for the efficiency optimization of a hybrid modulation for neutral-point-clamped dual-active-bridge (NPC-DAB) converter. The proposed D²EA approach realizes 99.92% efficiency modeling accuracy, and its feasibility is comprehensively validated in 2-kW hardware experiments, where the peak efficiency of 98.45% is attained. Overall, D²EA is data-light and can achieve highly accurate and highly practical data-driven models in one shot, and it is scalable to other applications, effortlessly.***

## I. INTRODUCTION

AS the public consciousness of global warming and the depletion of fossil fuel arises, renewable generation attracts widespread attention for a sustainable future. With the increasing penetration of renewable energy such as solar and wind, conventional power grid relies on power converters to interface with the distributed generation and regulate power transfer. The intermittency of renewable resources and the inherent nonlinearity of power converters pose serious threats to the power quality, grid stability, and aggravate the demand-supply mismatch [1].

Catering for the negative impacts induced by the more renewable generation in power grids, energy storage systems (ESSs) have been widely applied in renewable energy systems for reliable and flexible power management, like compressed air storage system, battery storage system, and flywheel energy system [2]. ESSs release the stored energy to the power grid during peak load hours and store the redundant power during off-peak periods to actively support the grid. The grid resiliency, power quality, stability, and reliability are enhanced with the utilization of ESSs.

The successful application of ESSs requires the bidirectional transfer capability of power converters, where the popular dual-active-bridge (DAB) converter is omnipresent by virtue of its high switching frequency, low power transmission loss, high power density, galvanic isolation, ease of zero switching, and good extendibility [3]. For example, in battery-based ESSs for residential application, a three-phase DAB dc-dc converter was applied for integrating photovoltaic panels, load shifting, and power backup [4]. Typically, a DAB converter is composed of two full semiconductor bridges, an in-between high-frequency transformer, and a leakage inductor, as proposed in 1990s [5].

Nevertheless, the conventional DAB converter is inappropriate for high-power ESS applications constrained by the power limits of semiconductor switches. The dilemma of the conventional DAB converter gives impetus to the developments of various DAB topologies. For instance, multilevel-dc-link DAB converters have been adopted for medium-voltage dc applications, as reviewed by Zhao et al. in [6]. Except for the multilevel topologies, cascading topologies such as the input-series output-parallel DAB converter have been utilized for high input voltage and high output current applications [7].

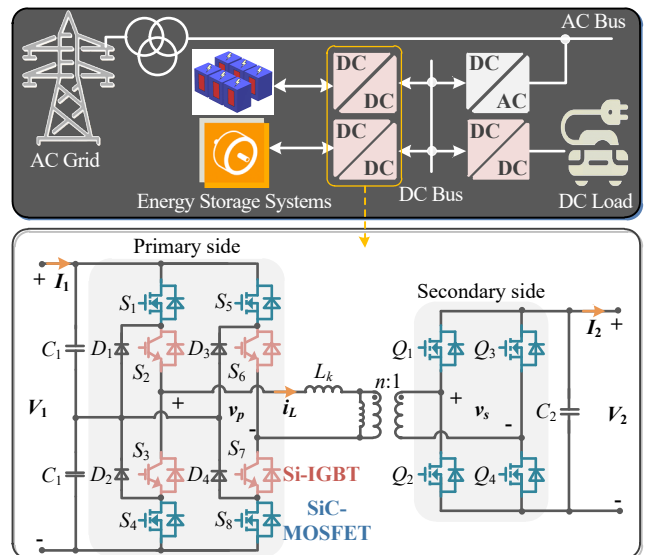

Fig. 1. NPC-DAB converters in ESSs.

In this paper, the three-level neutral-point-clamped DAB (NPC-DAB) converter adopted by Dong *et al.* [8] is the research target, which consists of two NPC bridge legs integrated by Si and SiC switches in the primary side, and a secondary SiC-based full bridge, as shown in Fig. 1. The hybrid switch choice of the NPC-DAB attains high efficiency while still maintaining cost effectiveness [8]. Aiming at enhancing the efficiency performance of the NPC-DAB converter under different circumstances, a hybrid modulation strategy with three degrees of control freedom is investigated to optimize the modulation parameters. As shown in Fig. 2, the hybrid modulation employs the duty ratio strategy in the primary NPC bridge and the phase shift strategy in the secondary bridge, including: the outer phase shift $D_0$ between $S_2$ and $Q_1$, the duty ratio $D_1$ of $S_1$, and the inner phase shift $D_2$ between $Q_1$ and $Q_4$. The switches in the primary bridge are regulated with variable duty ratio to reduce the turn-off loss [8].

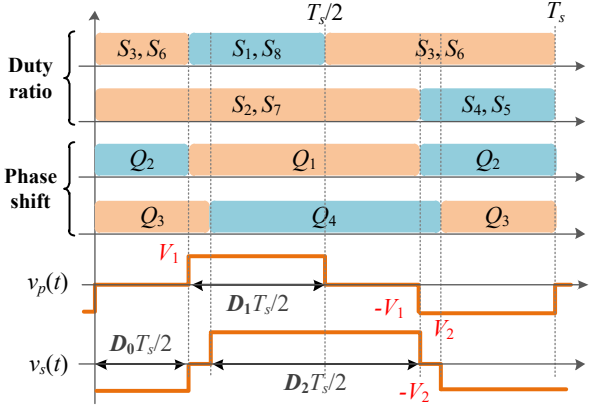

Fig. 2. The hybrid modulation for the NPC-DAB converter.

In this respect, the modulation objective should be modeled with the expectation of high modeling accuracy to precisely gauge the operating performance of DAB converters. Generally, the existing modeling approaches can be classified into two main categories, the conventional knowledge-based approach and the emerging data-driven approach.

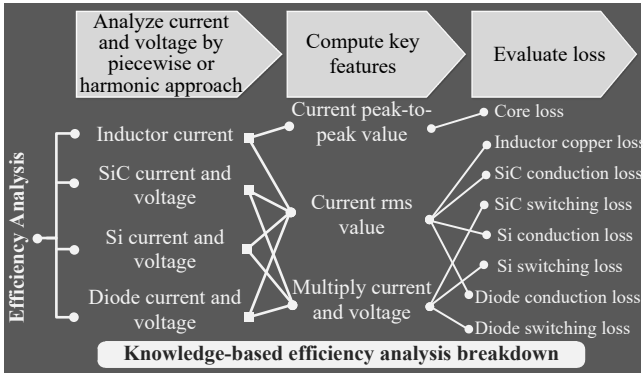

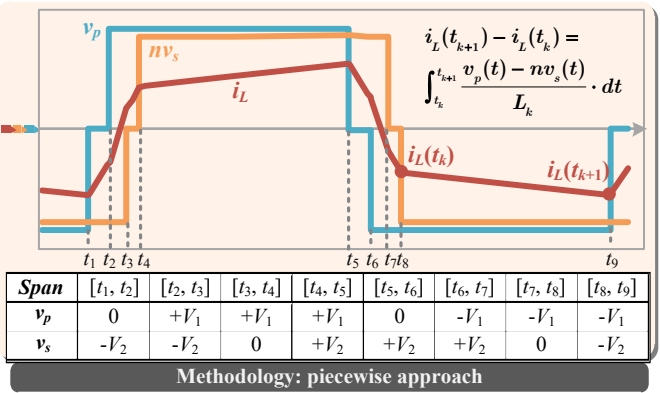

Fig. 3. The knowledge-based approach for the efficiency analysis.

The knowledge-based modeling method relies on expert knowledge for a thorough understanding of the converter working principles and the modulation behaviors, the human-dependent feature of which leads to low modeling accuracy and heavy manual burden. For example, the piecewise approach adopted in [9]–[11] deduced the ac inductor current segment-by-segment for all working modes, the high order of which indicates a time-consuming derivation process. The accuracy in [11], [12] was reduced due to the negligence of the inductor resistance and other parasitic parameters. In addition, the harmonic method for the efficiency analysis in [13], [14] smoothed out high-order harmonics to alleviate deduction complexity, but sacrificing the modeling accuracy.

The time-consuming efficiency analysis of the NPC-DAB converter under the hybrid modulation with knowledge-based approaches is given in Fig. 3. Initially, the current and voltage waveforms of the inductor and switches are analyzed with either piecewise approach or harmonic analysis, based on which some keys features are computed. Detailed losses are then evaluated [15]: the peak-to-peak inductor current evaluates magnetic core losses; inductor copper and device conduction losses are calculated with the current rms value; the voltage and current of devices are multiplied and averaged to evaluate switching losses. This derivation process should be repeated for all working modes, so is undoubtedly tedious considering the complex NPC-DAB topology and the intricate modulation. Moreover, the knowledge-based efficiency modeling can be inaccurate due to the inevitable model simplification and mathematical approximation.

As a step further, latest data-driven approaches for building the models of modulation objectives are the most advanced techniques. Compared to knowledge-based approaches, it can significantly relieve manpower burden, free engineers from tedious and repetitive works, and improve the modeling accuracy [16], [17]. To automatically analyze the current stress of DAB converters under triple phase shift (TPS) modulation, Li *et al.* trained neural networks (NNs) on the performance data obtained from simulation software [18]. In the efficiency-oriented optimization of TPS modulation, efficiency models have been built with NNs [19]. Besides, data-driven models for zero voltage switching (ZVS) conditions under the hybrid extended phase shift modulation were trained to optimize ZVS range [20]. Outlier detection algorithm has been integrated in the data-driven modeling of the current stress under a hybrid phase shift modulation [21]. The data-driven models in [18]–[21] can realize higher accuracy since the considerations of switching behavior and more precise circuit components contribute to more realistic simulation. These data-driven approaches, which achieve automation in modeling, are the future tendency to accelerate industrial design cycle.

However, as shown in Fig. 4, the existing data-driven methods suffer from two nontrivial drawbacks: data-intensive and model discrepancy. First, the commonly adopted data-driven methods such as neural networks for performance modeling are data-intensive, which require large datasets for training [16], [18]. However, it can be infeasible to acquire such large datasets in the power electronics domains. Second, the data-driven methods that are trained purely on simulation data suffer from model discrepancy, which induces nonnegligible deviations from the actual performance in real-world situations because of the following factors: unmodeled parasitic parameters, imprecise thermal modeling, disregarded physical phenomenon, deficient magnetic modeling, unforeseeable environmental disturbances, fluctuating ambient conditions, etc. For example, the effect of inductance fluctuations can cause divergence in the analysis of gain margin and phase margin in a cascaded power converter system [22]. [23], [24] discussed the negative influence of unmodeled parasitic resistance and capacitance on the state-space model accuracy. Besides, the nonlinearity of magnetic hysteresis loop, the skin and proximity effects under high switching frequency, and poor thermal modeling can bring about deviated loss modeling [25], [26].

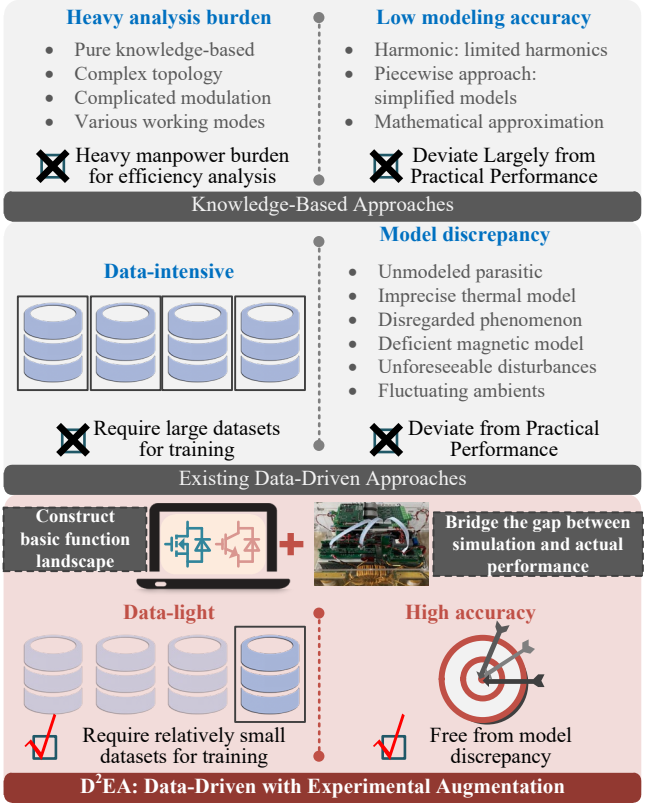

Fig. 4. Comparisons among the existing knowledge-based approaches, data-driven approaches, and the proposed D²EA modeling approach.

Aiming at the shortcomings of the existing approaches, a data-driven modeling with experimental augmentation (D²EA) is proposed for performance modeling and optimization. The essential idea of D²EA is to combine both simulation and experimental results in the data-driven modeling. It is applied in the efficiency-oriented optimization of the NPC-DAB converter under the hybrid modulation. Generally, D²EA includes three main steps. In Step 1, simulation is built, and hardware experiments are conducted to collect sufficient simulation data and experimental results. Provided the hybrid data pool, Step 2 adopts an ensemble learning algorithm, the extreme gradient boosting (XGBoost) algorithm, to train accurate data-driven models for efficiency. In Step 3, the latest particle swarm optimization with state-based adaptive velocity limit (PSO-SAVL) algorithm is applied to find the best modulation parameters that achieve the optimal efficiency. In D²EA modeling, the simulation data obtained from simulation software and experimental data collected in practice form a hybrid data pool, both of which are indispensable for the data-light and accurate performance modeling. First, the simulation data constructs the basic functional landscape of the modulation objectives, which serves as a time-efficient and cost-effective data source to reduce the amount of experimental data required. Furthermore, the experimental data bridges the gap between the simulation data and real-life implementation, enhancing the practicality of the proposed D²EA approach. The comparisons of different modeling approaches are summarized

in Table I and Fig. 4, and the major contributions of this paper are concluded as follows:

- The proposed D²EA promotes the existing data-driven approaches for performance modelling in two aspects: First, the D²EA approach is data-light, since both data sources are contributing to the performance modelling, and the adopted XGBoost algorithm requires less training data compared with other data-driven algorithms; Second, the D²EA approach can achieve accurate and practical modeling in one shot, mitigating the needs for repetitive tuning of simulation and data-driven models to match practical situations.

- In terms of the contribution for the application background, with the D²EA approach, the complicated efficiency modeling is automated and high accuracy can be realized. Besides, this paper initiates the efficiency optimization for the NPC-DAB converter under the hybrid modulation.

TABLE I
COMPARISONS OF DIFFERENT MODELING APPROACHES

| CATEGORY | REFERENCE | MANPOWER BURDEN | DATA SIZE | DEVIATION FROM PRACTICE |
|---|---|---|---|---|
| Knowledge-based: Piecewise | [9]–[11], [27] | Large | - | Large |
| Knowledge-based: Harmonics | [12]–[14], [28] | Medium | - | Large |
| Existing Data-Driven | [18]–[21] | Low | Data-intensive | Medium |
| **Proposed D²EA** | **This Paper** | **Low** | **Data-light** | **Low** |

The organization of this paper is as follows. Section II elaborates the methodologies of the proposed D²EA modeling in detail. Section III studies a design case in a comprehensive way. Hardware experiments presented in Section IV verify the feasibility of the proposed D²EA. In Section V, the conclusion of the paper is summarized.

## II. METHODOLOGY OF THE PROPOSED D²EA MODELING

In this section, the methodology of the proposed D²EA modeling is comprehensively illustrated. D²EA is a universal modeling approach which can be extensively applied to various converter circuits, modulation strategies, and applications. For illustrative purposes, the proposed D²EA approach is applied to shed lights on the efficiency-oriented optimization for the NPC-DAB converter under the hybrid modulation.

### A. The Proposed D²EA Modeling Approach

The design process is shown in Fig. 5, which consists of three main steps. First, simulation models are run and hardware experiments are carried out to collect sufficient simulation data and experimental performance, respectively. Second, utilizing the hybrid data pool of simulation data and experimental data, two XGBoost-based data-driven models are consecutively trained to deliver accurate evaluation of actual efficiency performance in the physical world. In Step 2, one initial XGBoost model constructs a basic functional landscape, and a subsequent XGBoost model bridges the accuracy gap to compensate for the simulation model discrepancy. Third, the PSO-SAVL algorithm interacts with the trained XGBoost models to achieve the optimal efficiency over the entire working ranges.

Generally, the proposed D²EA modeling is implemented in a high level of design automation, freeing human experts from complicated, time-consuming, and error-prone process of performance derivation and parameter optimization.

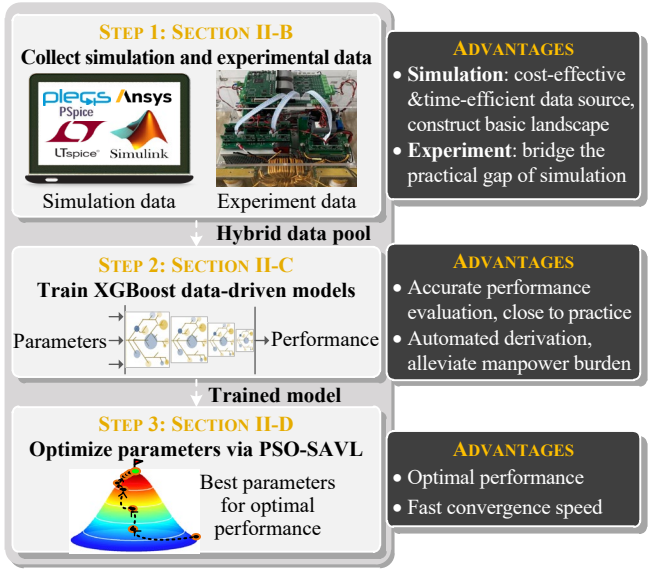

Fig. 5. Flowchart of the proposed D²EA modeling approach.

## B. Step 1: Collect Simulation and Experimental Data

As a preliminary preparation, the design specifications such as the performance to optimize, parameters to consider, and working conditions are determined. In this paper, the values of duty ratio ($D_1$) and inner phase shift ($D_2$) should be designed under various load conditions ($P_L$) to optimize efficiency. Hence, the parameters incorporate $D_1$, $D_2$, and $P_L$, and the modulation objective is the optimal efficiency.

Thereafter, in Step 1, simulation models are built and run for sufficient number of times to collect performance data. The simulation data is responsible for establishing the basic functional relationships between design parameters and performance.

To mitigate the negative impact brought by mismatched simulation models, experimental data is collected in real-world situations, which is used in the D²EA data-driven modeling for approaching practical performance. The experimental data contributes to a more complete data pool.

Both simulation and experimental data are necessary to guarantee the satisfactory modeling accuracy. Simulation results reduce the number of experiments required through generating a basic functional landscape, fostering the data-light merit of the proposed D²EA approach. Experimental data further improves the modeling accuracy. Another merit of D²EA is that the repetitious tuning of models to match practical situations in the existing approaches based on pure simulation is avoided, and promising data-driven models with high accuracy and high practicality can be trained in one shot. After Step 1 of the proposed D²EA modeling, a hybrid data pool is generated.

## C. Step 2: Train Data-Driven Models of Efficiency via the XGBoost Algorithm

Step 2 of the proposed D²EA modeling automatically learns the functional relationships between the considered parameters and the modulation objective through the XGBoost algorithm. XGBoost is a popular ensemble learning algorithm [29], which utilizes a set of weak decision trees to learn the target. In the power electronics domain of performance modeling in this paper, XGBoost algorithm is chosen because of its easily scalable structure, fast computation speed, simple implementation, etc. First, XGBoost model is flexible structure-wise, and its learning capacity can increase by simply stacking more base models, which is more powerful than conventional machine learning algorithms such as support vector machine [20]. Second, XGBoost model can be parallelly applied during the inference stage, justifying its fast inference speed. Furthermore, XGBoost algorithm is simple to implement since it has few hyperparameters to adjust to find the optimal structure.

To attain high accuracy and high practicality and coordinate with the hybrid data pool, two XGBoost models are consecutively trained. XGBoost-I model forms landscape basis, and XGBoost-II model bridges the accuracy gap between simulation and experiments to avoid model discrepancy.

The training of the XGBoost models is accomplished with the sequential gradient boosting strategy shown in Fig. 6, where there are $L_1$ and $L_2$ decision trees in XGBoost-I and XGBoost-II, respectively. For the training of XGBoost-I, the subsequent decision tree is trained on the residual $o_{l,sim}^*$, as expressed in (1) and (2). $o_{l,sim}^*$ represents the difference between the ground-truth objective $o_{sim}^*$ (simulation efficiency $\eta_{sim}$) and the aggregate sum of all the previous decision trees from 1 to $l$-1. For the $l^{th}$ tree, it receives $o_{l-1,sim}^*$ to calculate the $l^{th}$ residual $o_{l,sim}^*$, and its adjustable weights $\theta_{l,sim}$ are tuned to minimize the difference between $o_{l,sim}^*$ and the output $y_{l,sim}$, as shown in Fig. 5 and (1). Afterwards, the resulting error between $o_{l,sim}^*$ and $y_{l,sim}$ is provided to the next $(l+1)^{th}$ tree for training. This recurrent boosting-based training process is repeated for $L_1$ decision trees. Thereafter, the XGBoost-II model is similarly trained as formularized in (3) and (4), where the ground-truth objective $o_{gap}^*$ denotes the gap between experimental efficiency $\eta_{exp}$ and the results from XGBoost-I to mitigate model discrepancy.

Besides, the performance evaluation utilizes the trained XGBoost-I and XGBoost-II models to infer the efficiency of the given input parameters ($D_1$, $D_2$, $P_L$), as expressed in (5).

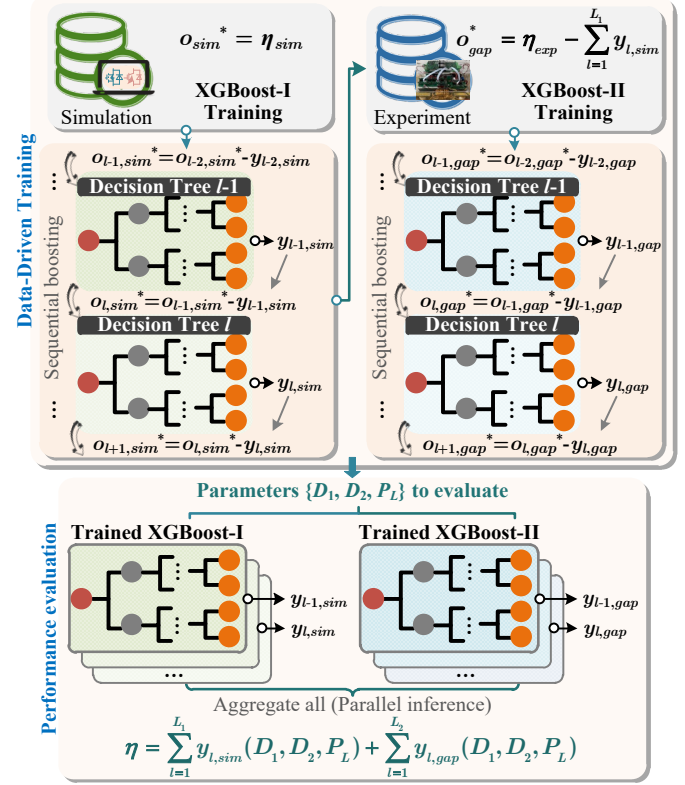

Fig. 6. Training and inference of the XGBoost algorithm.

In this paper, the inputs of XGBoost models include $D_1$, $D_2$, and $P_L$, which can be extended to consider other impactful parameters, and the key point is to acquire training data which can properly represent the considered application scenarios. As shown in Fig. 7, the input variables of the D²EA modeling can be extended to incorporate modulation parameters, operating conditions, and circuit parameters. For instance, the effects of output power $P_L$, voltage conversion ratio $m$, and the fluctuation of circuit parameters can be considered in the efficiency modeling.

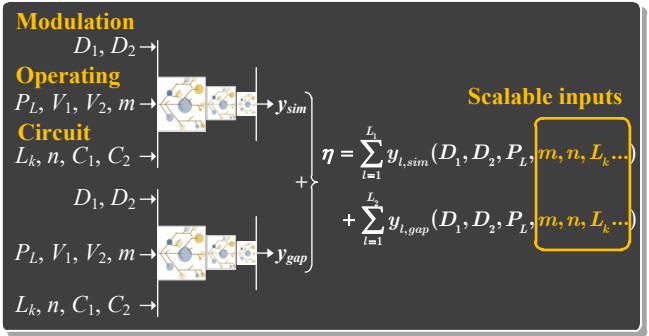

Fig. 7. XGBoost models considering other impactful parameters.

With Step 2 of the proposed D²EA modeling, data-driven XGBoost models are properly trained, serving as the surrogate models for the efficiency of the NPC-DAB converter under the hybrid modulation. Being informed of both simulation data and experimental performance in practice, the XGBoost models can provide accurate and practical efficiency evaluation, which improves the utility of the proposed D²EA approach.

$$\min_{\theta_{i,sim}} (obj_{i,sim}) = \min_{\theta_{i,sim}} (o^*_{i,sim} - y_{i,sim})^2, l=1,...,L_1 \quad (1)$$

$$o^*_{l,sim} = o^*_{l-1,sim} - y_{l-1,sim} = o^*_{sim} - \sum_{i=1}^{l-1} y_{i,sim} = \eta_{sim} - \sum_{i=1}^{l-1} y_{i,sim} \quad (2)$$

$$\min_{\theta_{i,gap}} (obj_{i,gap}) = \min_{\theta_{i,gap}} (o^*_{i,gap} - y_{i,gap})^2, l=1,...,L_2 \quad (3)$$

$$o^*_{l,gap} = o^*_{l-1,gap} - y_{l-1,gap} = o^*_{gap} - \sum_{i=1}^{l-1} y_{i,gap} = \eta_{exp} - \sum_{i=1}^{L_1} y_{i,sim} - \sum_{i=1}^{l-1} y_{i,gap} \quad (4)$$

$$\eta = y_{sim} + y_{gap} = \sum_{l=1}^{L_1} y_{i,sim}(D_1, D_2, P_L) + \sum_{l=1}^{L_2} y_{i,gap}(D_1, D_2, P_L) \quad (5)$$

### D. Step 3: Search for Optimal Modulation Parameters via the PSO-SAVL Algorithm

$$f_e = \frac{d_g - d_{min}}{d_{max} - d_{min}} \quad (6)$$

$$d_i = \frac{1}{N-1} \sum_{j=1, j \neq i}^{N} \sqrt{(D_{1,i} - D_{1,j})^2 + (D_{2,i} - D_{2,j})^2} \quad (7)$$

$$VL = \frac{1}{1 + \left(\frac{1}{vl_{max}} - 1\right) \exp\left(\ln\left(\left(\frac{1}{vl_{max}} - 1\right) \middle/ \left(\frac{1}{vl_{min}} - 1\right)\right) f_e\right)} \cdot X_{max} \quad (8)$$

$$V_i = \omega V'_i + c_1 r_1 (pb_i - X_i) + c_2 r_2 (gb - X_i) \quad (9)$$

In Step 3, a latest PSO variant, PSO-SAVL, is chosen to search for the optimal modulation parameters to reach the best efficiency over the entire operating ranges. Compared with the conventional PSO algorithm, PSO-SAVL adaptively adjusts the velocity limit of particles to match the evolutionary state of population, which improves the global exploration capability and the convergence speed. The superiority of the adopted PSO-SAVL is theoretically and empirically validated in [30].

The flowchart of PSO-SAVL is shown in Fig. 8. In the beginning, the load conditions $P_L$ and the algorithm hyperparameters including particle number $N$, velocity inertia

$\omega$, velocity limit coefficients $vl_{max}$, $vl_{min}$, and learning factors $c_1$, $c_2$ are initialized. Position $X_i$ (i.e., $D_{1,i}$, $D_{2,i}$) and velocity $V_i$ (i.e., $V_{1,i}$, $V_{2,i}$) are uniformly initialized. Practical efficiency performance $\eta$ in physical world is evaluated through interfacing the trained XGBoost models from Step 2, and the global best position $gb$ and historical best position $pb_i$ are updated. Hereafter, $f_e$ which measures the evolutionary state of particles is computed to tune the velocity limit $VL$ (i.e., $VL_1$, $VL_2$), as shown in (6), (7), and (8), where $X_{max}$ is the position boundary, $d_g$ is the $d_i$ value of the best particle, and $d_{min}$, $d_{max}$ are the minimum and maximum $d_i$ values. The velocity $V_i$ of each particle is then adjusted with (9) and restricted by the new $VL$, and the adjusted $V_i$ is used next for relocating the particles. This process repeats until the maximum iteration is met.

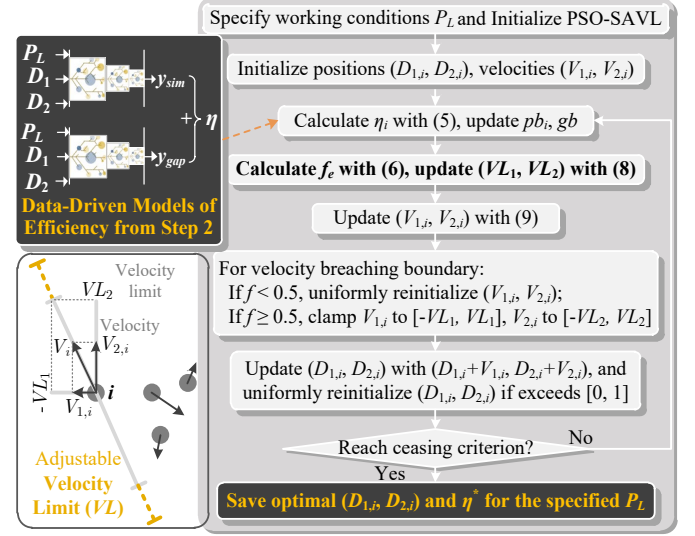

Fig. 8. Flowchart of the PSO-SAVL algorithm for optimizing modulation parameters to achieve the best efficiency.

In summary, with the assistance of PSO-SAVL algorithm, Step 3 finds the optimal modulation parameters to achieve the best efficiency over the entire load range.

### E. Strategies to Tune the Hyperparameters of XGBoost Models and the PSO-SAVL Algorithm

The strategies and insights in selecting the hyperparameters of the XGBoost models and the PSO-SAVL algorithm are revealed. To tune the hyperparameters of XGBoost for high modeling accuracy, the collected data is partitioned into training, test, and validation sets. The potential hyperparameter values are enumerated and the ones with the highest accuracy on the test set are selected as the optimal hyperparameters. In terms of PSO-SAVL algorithm, the hyperparameters discussed in Section II-D are adjusted to achieve a proper balance between global exploration and local exploitation [30].

## III. Design Case with the Proposed D²EA Modeling

Section III discusses a design case given by the proposed D²EA modeling approach in a step-by-step manner. Section III-A, III-B, and III-C elaborate the settings and results of Step 1, Step 2, and Step 3, respectively. Section III-D sheds light on the computational costs required for the D²EA modeling.

### A. Step 1: Collect Simulation and Experimental Data

Design specifications that provide operating boundaries for simulation and experiments are summarized in Table II. In this paper, considering the application of ESSs shown in Fig. 1, the input and output voltages are fixed, which regulated by the ESSs and DC bus, respectively. Other design specification such as the switching frequency is not limited to the selected value 20 kHz, and other values can be adopted with slight modifications in the initial data acquisition step.

TABLE II
DESIGN SPECIFICATIONS

| WORKING CONDITIONS | | | |
|---|---|---|---|
| Input voltage $V_{1R}$ | 300 V | Output voltage $V_{2R}$ | 140 V |
| Rated power $P_{LR}$ | 2 kW | Frequency $f_s$ | 20 kHz |
| SEMICONDUCTOR SWITCHES | | | |
| Si-IGBT | IKW40N65ES5 | SiC-MOSFET | UF3C065030K4S |
| Clamping diode | APT30DQ60BG | Dead time | 400 ns |
| MAGNETICS | | | |
| Duty ratio n:1 | 2:1 | External inductance $L_k$ | 236 μH |
| RANGES OF PARAMETERS | | | |
| Primary duty ratio $D_1$ | [0, 1] | Inner phase shift $D_2$ | [0, 1] |
| Outer phase shift $D_0$ | [0, 1] | Output power $P_L$ | [200 W, 2 kW] |

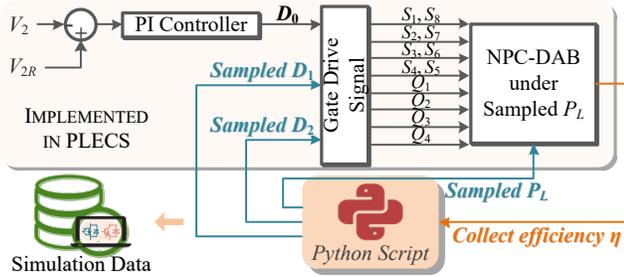

Fig. 9. Closed-loop simulation conducted by python script for all sampled $D_1$, $D_2$, and $P_L$.

In Step 1, a closed-loop simulation model is built in PLECS, as shown in Fig. 9. The output of the PI controller is the outer phase shift $D_0$ between the primary side and the secondary side, which regulates the power transfer. To generate sufficient simulation data, within the ranges of three parameters $D_1$, $D_2$, and $P_L$, 25, 25, and 20 number of samples are evenly selected, respectively. Hence, the PLECS simulation is repetitively run for the chosen parameter values to collect 12,500 simulation data automatically using python scripts.

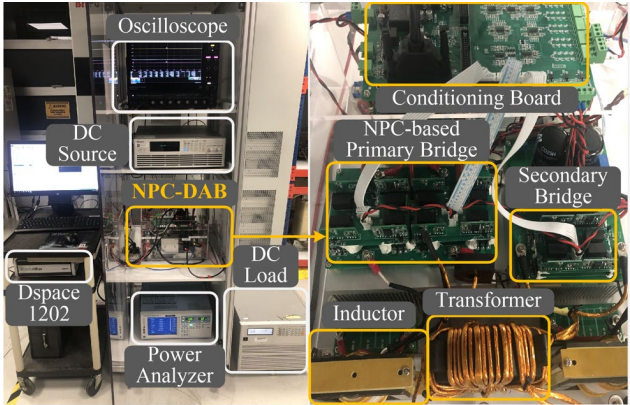

Fig. 10. Experimental platform and hardware prototype.

During the automated running of simulation, hardware experiments are conducted to collect practical efficiency performance to mitigate the deviation between simulation and practice. The experimental platform and the built hardware prototype are shown in Fig. 10, where the equipment includes a Chroma dc power supply and an ac/dc programmable load, a Dspace 1202, a Lecroy oscilloscope, and an NPC-DAB converter. To measure the efficiency in experiments, a Yokogawa WT3000 precision power analyzer is used. In the design case, there are totally 1000 experimental data collected for various values of $D_1$, $D_2$, and $P_L$.

### B. Step 2: Train Data-Driven Models of Efficiency via the XGBoost Algorithm

Given the hybrid pool of simulation data and experimental data, Step 2 of the proposed D²EA approach trains XGBoost models to learn the underlying relationships between the design parameters ($D_1$, $D_2$, $P$) and the efficiency $\eta$ to be optimized.

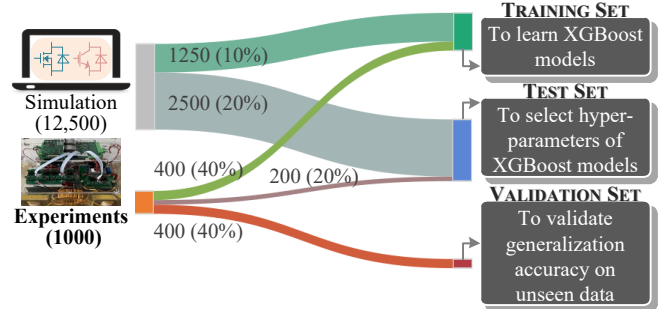

Fig. 11. Partitions of simulation and experimental data.

TABLE III
SETTINGS OF EXTREME GRADIENT BOOSTING MODELS

| | TRAINING SET | TEST SET | VALIDATION SET |
|---|---|---|---|
| Simulation data | 1250 (10%) | 2500 (20%) | 8750 (70%) |
| Experimental data | 400 (40%) | 200 (20%) | 400 (40%) |
| XGBOOST SPECIFICATIONS | | | |
| Inputs | $D_1, D_2, P$ | Outputs | Efficiency $\eta$ |
| HYPERPARAMETERS OF XGBOOST-I | | | |
| Max tree height | 11 | Number of trees | 140 |
| L2 regularization | 1.0 | Learning rate | 0.05 |
| HYPERPARAMETERS OF XGBOOST-II | | | |
| Max tree height | 9 | Number of trees | 94 |
| L2 regularization | 0.01 | Learning rate | 0.1 |

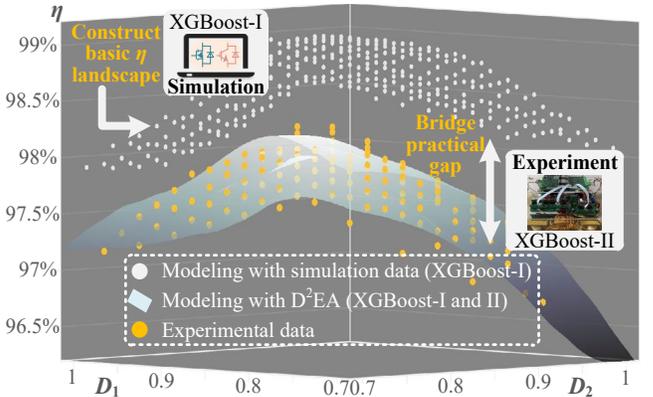

Fig. 12. Modeling with D²EA under 1 kW output power.

By following the training process in Fig. 6, XGBoost surrogate models for efficiency are automatically trained, the settings of which are summarized in Table III. The collected 12,500 simulation data and 1000 experimental data are partitioned into three datasets for training the XGBoost models, refining modeling accuracy, and validating the generalization capability, respectively, as displayed in Fig. 11.

The modeling results of Step 2 are intuitively shown in Fig. 12, verifying the utility of both simulation and experimental data. In Fig. 12, the yellow scattered dots are the experimental results captured in hardware experiments, the white scatted dots are the efficiency predictions of XGBoost-I (trained purely on simulation data), and the transparent surface indicates the accurate efficiency modeling of D²EA, which combines both XGBoost-I and XGBoost-II. As can be seen, XGBoost-I constructs a basic functional landscape of the efficiency, the shape and trend of which are similar to the actual performance behavior. However, XGBoost-I suffers from model discrepancy, as the average difference between the modeling with only simulation data and the experimental data is around 1.1%, which is a nontrivial gap. XGBoost-II, which is trained on experimental data, significantly bridges the gap between the simulation results and the experimental performance. With both XGBoost-I and XGBoost-II, the data-driven models with D²EA can precisely evaluate the practical efficiency of the NPC-DAB converter in the physical world.

Furthermore, to quantitatively validate the high accuracy and practicality of the proposed D²EA, Fig. 13 compares the modeling accuracy of three approaches on experimental data. If only simulation data is used for modeling, the average percentage difference is more than 1%, which is a nonnegligible gap for efficiency modeling. Although the efficiency modeling with only experimental data can decrease the error to 0.58%, the D²EA approach attains 99.92% average percentage accuracy on validation dataset by leveraging both simulation and experimental data.

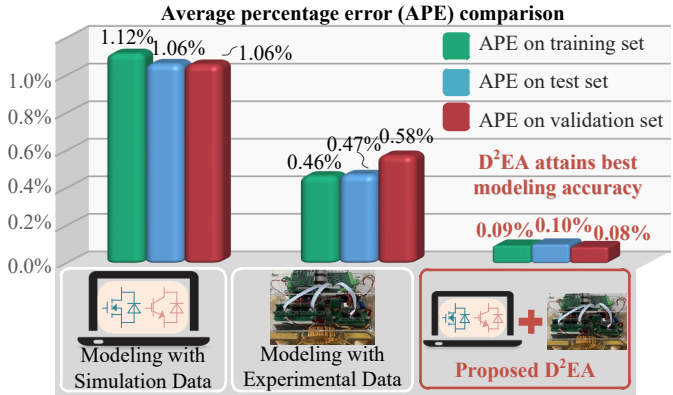

Fig. 13. Accuracy comparison (on experimental data) among three modeling approaches: Modeling with only simulation data, modeling with only experimental data, and modeling with the proposed D²EA.

### C. Step 3: Search for Optimal Modulation Parameters via the PSO-SAVL Algorithm

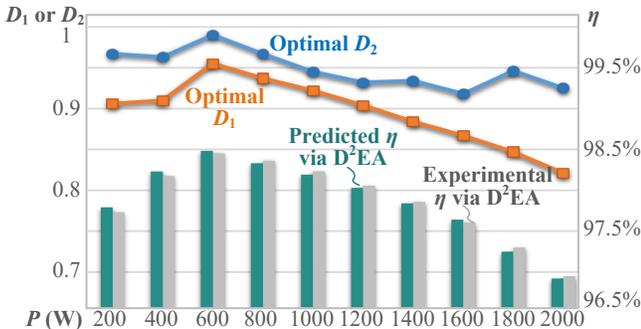

Fig. 14. Optimal $D_1$ and $D_2$ for the entire load range and accuracy comparison on optimal points.

Utilizing the PSO-SAVL algorithm in Fig. 8, Step 3 finds the optimal values of $D_1$ and $D_2$ to achieve the best efficiency for various load conditions $P$. Configurations of the adopted PSO-SAVL are summarized in Table IV. The outcomes of Step 3 are shown in Fig. 14, based on which $D_1$ and $D_2$ are adjusted in real time to achieve optimal efficiency under varying load conditions. The difference between the predicted efficiency via D²EA and the experimental efficiency in physical world is trivial, validating the high modeling accuracy and high practicality of the proposed D²EA modeling approach.

| | |
|---|---|
| Inputs | Trained data-driven surrogate models XGBoost-I $y_{sim}(D_1, D_2, P)$ and XGBoost-II $y_{exp}(D_1, D_2, P)$ from Step 2 |
| Outputs | Optimal $D_1$, $D_2$ and the highest efficiency $\eta^*$ for given $P$ |
| Population size | 10 |
| Total number of iterations | 50 |
| Velocity limit coefficients | $vl_{min} = 0.05$; $vl_{max} = 0.2$ |
| Learning factors | $c_1 = c_2 = 2.05$ |
| Weight inertia factor | $\omega$ decreases from 0.9 to 0.1 |

### D. Computational Costs of the Proposed D²EA Modeling

TABLE V
COMPUTATIONAL COSTS OF EACH STEP OF D²EA

| MAIN STEP | AVERAGE CPU TIME | CPU UTILIZATION |
|---|---|---|
| Step 1: Collect simulation and experimental data | 9 hours and 17 minutes | 20.2% |
| Step 2: Train data-driven models of efficiency | 1 minute 54.7 seconds | 68.9% |
| Step 3: Search for Optimal Modulation Parameters | 12 minutes 25 seconds | 43.9% |

One major merit of the proposed D²EA approach is that the model building process and modulation optimization are automated with computer platforms, which greatly alleviates human involvements and largely accelerates design cycle. To disclose the computational costs, the average implementation time and CPU utilization of each step of D²EA are recorded in Table V, where the workstation configures a 4-core Intel Xeon processor E5-1630 with 16 gigabytes RAM. The collection of simulation and experimental data occupies the majority of implementation time, while the training of data-driven models exhibits the highest average CPU utilization.

### E. Comparison of Modeling Accuracy

The data-light merit of the proposed D²EA approach is empirically verified. The comparison results between D²EA and other popular data-driven algorithms are summarized in Fig. 15. All the compared algorithms are run for 30 times on experimental data, where the dotted lines represent the average modeling accuracy of different algorithms, and the areas denote the variation of modeling accuracy. From Fig. 15, the D²EA approach achieves the highest and the most robust modeling accuracy in all data sizes, and the accuracy reaches 99.88% with only 10% of data, justifying the data-light merit.

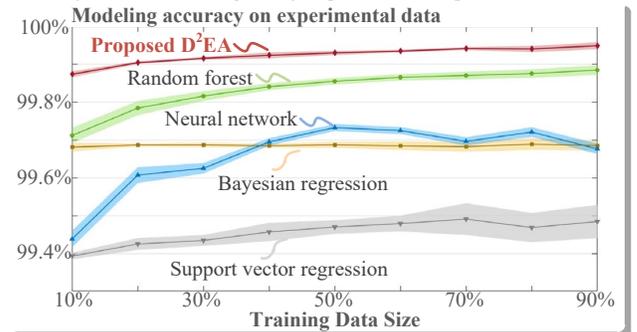

Fig. 15. Comparison of the modeling accuracy on experimental data among the proposed D²EA approach, SVR, BR, RF, and NN.

## IV. DESIGN CASE WITH THE PROPOSED D²EA MODELING

This section holistically presents hardware experimental results to verify the feasibility of the proposed D²EA modeling approach. The operating conditions are given in Table II, and the hardware platform is shown in Fig. 10.

### A. Steady-State Waveforms

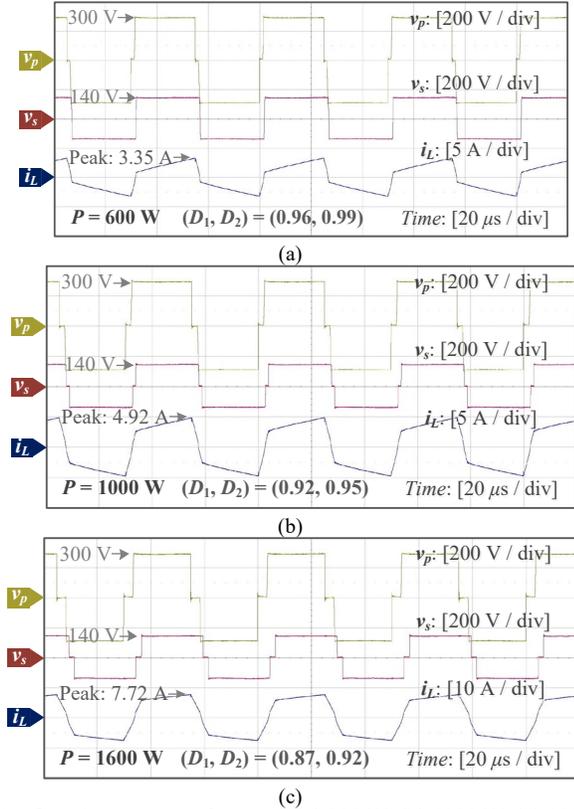

Fig. 15. Steady-state waveforms of: (a) 600 W, (b) 1000 W, (c) 1600 W.

To comprehensively validate the steady-state operations over the entire load range, 600 W (30% load), 1000 W (50% load), and 1600 W (80% load) are analyzed as examples, which represent light, medium, and heavy load levels, respectively. As high-frequency ac voltages $v_p$, $v_s$ and current $i_L$ shown in Fig. 15, $D_1$ and $D_2$ values are adjusted to the optimal values in Fig. 14 to reduce reverse current and improve efficiency by introducing adequate amount of zero-level voltage plateaus.

### B. Dynamic Response of Load Step

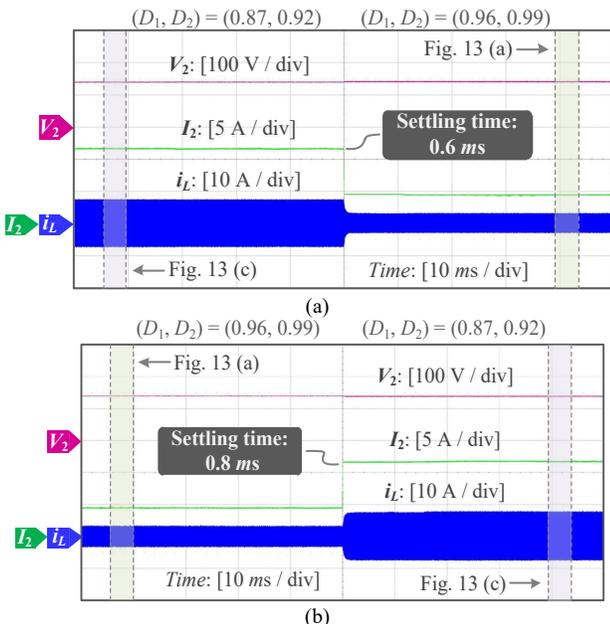

Fig. 16. Dynamic response waveforms when: (a) $P$ changes from 1600 W to 600 W, (b) $P$ changes from 600 W to 1600 W.

Fig. 14 shows the dynamic response of the optimal hybrid modulation when load steps are injected. The output voltage is robustly controlled, and the transient settling time is less than 1 $ms$. The zoom-in views when $P = 600$ W and $P = 1600$ W are displayed in Figs. 13 (a) and (c), respectively.

### C. Comparison of Efficiency Performance

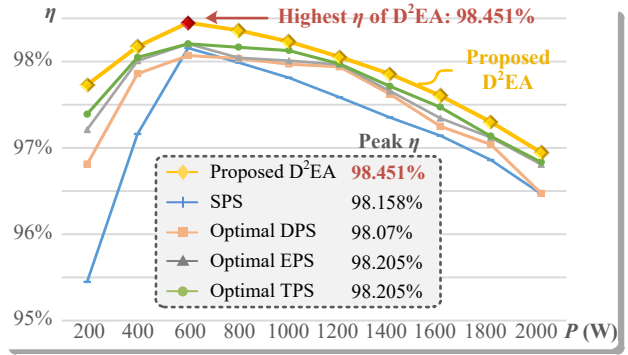

Fig. 17. Efficiency comparisons among the D²EA, SPS, optimal DPS, optimal DPS, optimal EPS, and optimal TPS.

The superiority of the proposed D²EA approach compared with other modulation strategies is experimentally validated, as shown in Fig. 17. The proposed D²EA approach realizes the highest efficiency 98.45% at $P = 600$ W, and it is significantly better than single phase shift (SPS) and optimal dual phase shift (DPS) modulation by 0.293% and 0.381%, respectively. Compared with the advanced strategies, optimal extended phase shift (EPS) and optimal triple phase shift (TPS) modulation, D²EA achieves higher efficiency consistently, and the peak efficiency is improved by 0.246%.

To demonstrate the high efficiency obtained with the D²EA approach, detailed loss breakdown charts are provided in Fig. 18. Under light load conditions of $P = 600$ W, the switching and conduction losses of the secondary bridge are the main losses. With the increasing power level, the percentage of the losses of secondary bridge decreases, while that of inductor and transformer losses increases. The percentage of the losses of the primary NPC bridge is less than that of the secondary bridge.

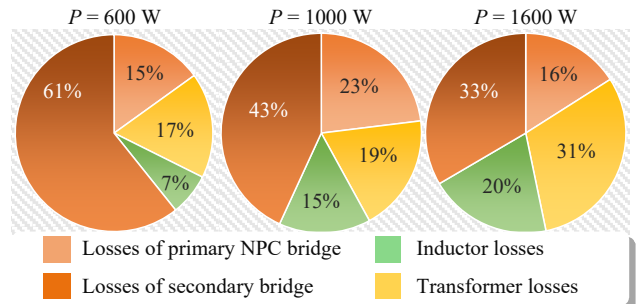

Fig. 18. Loss breakdown when load $P$ is 600 W, 1000 W, and 1600 W.

Besides, benefiting from the high accuracy and practicality of the D²EA modeling, the optimized modulation values $D_1$ and $D_2$ from D²EA will realize better efficiency compared with those from the modeling with simulation data only and the modeling with experimental data only. Fig. 19 presents the comparison results, where D²EA attains higher efficiency than the optimized hybrid modulation with simulation-based modeling and experiment-based modeling by 0.243% and 0.236%, respectively.

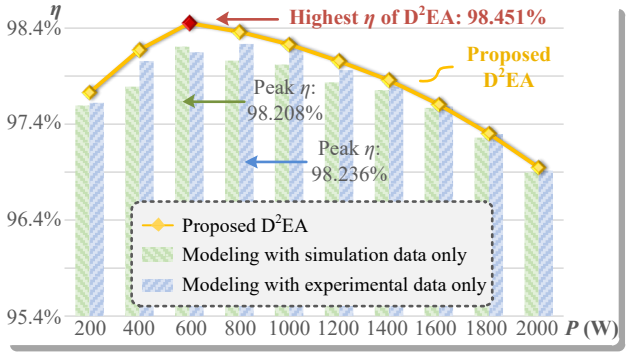

Fig. 19. Efficiency comparisons among the D²EA, the optimized hybrid modulation via simulation-based modeling, and the optimized hybrid modulation via experiment-based modeling.

### D. Verification of Optimality

Aiming at proving the optimality of the modulation values obtained from the proposed D²EA approach, the experimental efficiency results of adjacent $D_1$ and $D_2$ values near the optimized values are measured, as shown in Fig. 20. Under various load conditions, D²EA can precisely locate the optimal efficiency areas, validating the optimality.

With the comprehensive experiments in Section IV, the feasibility of the proposed D²EA method is verified.

## V. CONCLUSION

Striving to conquer the challenges in the mainstream knowledge-based and latest data-driven approaches for performance modeling, this paper has proposed a novel data-driven modeling with experimental augmentation (D²EA). In D²EA, the simulation data serves as cost-effective and time-efficient data source to construct basic functional landscape. While the augmentation of experimental data contributes to bridging the gap between the data-driven surrogate models and practical performance in physical world, alleviating model mismatch. In comparison with the existing data-driven approaches, the proposed D²EA approach is data-light, and the built performance models are highly accurate and practical, free from model discrepancy.

The application of D²EA focuses on the efficiency-oriented optimization of a hybrid modulation for NPC-DAB converters in energy storage systems. D²EA can be effortlessly extended to other application scenarios. Overall, it consists of three steps. In Step 1, simulation is automatically run and hardware experiments are conducted to collect both simulation and experimental data. In Step 2, the advanced extreme gradient boosting algorithm trains accurate data-driven models for the design performance. Two XGBoost data-driven models are consecutively trained by leveraging residuals to produce practical efficiency evaluation. In Step 3, a latest PSO variant searches for the optimal modulation parameters to realize the best efficiency over the entire load range.

Experiments on a 2-kW hardware prototype comprehensively validate the high modeling accuracy and high practicality and thus prove the effectiveness of the proposed D²EA modeling method. In the design case, via the proposed D²EA approach, the modeling accuracy for efficiency achieves 99.92%. Moreover, the efficiency performance optimized via the D²EA is consistently better than other advanced modulation strategies, and the highest efficiency reaches 98.45%.

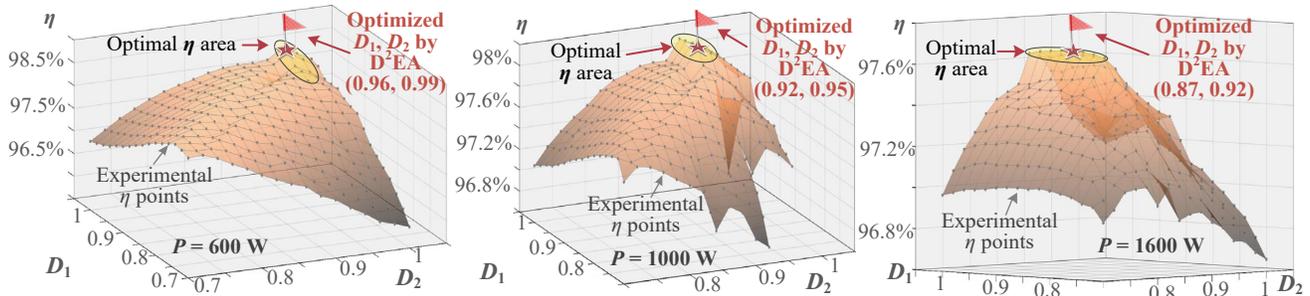

Fig. 20. Validation of optimality under 600 W, 1000 W, and 1600 W.

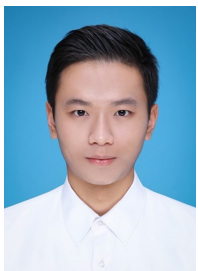

**Xinze Li** received his bachelor's degree in Electrical Engineering and its Automation from Shandong University, China, 2018. He has been awarded the Ph.D. degree in Electrical and Electronic Engineering from Nanyang Technological University, Singapore, 2023.

His research interests include dc-dc converter, modulation design, digital twins for power electronics systems, design process automation, light and explainable AI for power electronics with physics-informed systems, application of AI in power electronics, and deep learning and machine learning algorithms.

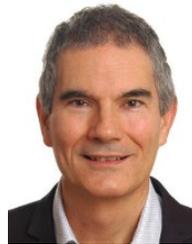

**Josep Pou (Fellow, IEEE)** received the B.S., M.S., and Ph.D. degrees in electrical engineering from the Technical University of Catalonia (UPC)-Barcelona Tech, in 1989, 1996, and 2002, respectively.

In 1990, he joined the faculty of UPC as an Assistant Professor, where he became an Associate Professor in 1993. From February 2013 to August 2016, he was a Professor with the University of New South Wales (UNSW), Sydney, Australia. He is currently a Professor with the Nanyang Technological University (NTU), Singapore, where he is Cluster Director of Power Electronics at the Energy Research Institute at NTU (ERI@N) and co-Director of the Rolls-Royce at NTU Corporate Lab. From February 2001 to January 2002, and February 2005 to January 2006, he was a Researcher at the Center for Power Electronics Systems, Virginia Tech, Blacksburg. From January 2012 to January 2013, he was a Visiting Professor at the Australian Energy Research Institute, UNSW, Sydney. He has authored more than 430 published technical papers and has been involved in several industrial projects and educational programs in the fields of power electronics and systems. His research interests include modulation and control of power converters, multilevel converters, renewable energy, energy storage, power quality, HVdc transmission systems, and more-electrical aircraft and vessels.

He is Associate Editor of the IEEE Journal of Emerging and Selected Topics in Power Electronics. He was co-Editor-in-Chief and Associate Editor of the IEEE Transactions on Industrial Electronics. He received the 2018 IEEE Bimal Bose Award for Industrial Electronics Applications in Energy Systems.

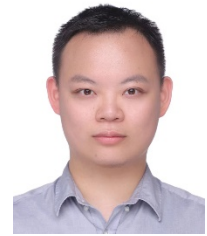

**Jiaxin Dong** received the B.S. degree in electrical engineering from Southeast University, China, in 2016. He is currently working toward the Ph.D. degree in the School of Electrical and Electronic Engineering, Nanyang Technological University, Singapore. His research interests include modulation and control of power electronics, multilevel converters, and energy storage systems.

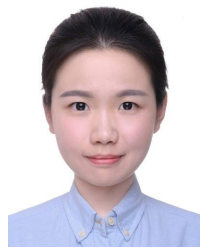

**Fanfan Lin** was born in Fujian, China in 1996. She received her bachelor's degree in electrical engineering from Harbin Institute of Technology in China in 2018. She received her joint Ph.D. degree in Nanyang Technological University, Singapore, and Denmark Technical University, Denmark in 2022. Her research interest includes the deep learning, domain adaptation, power converter design with artificial intelligence.

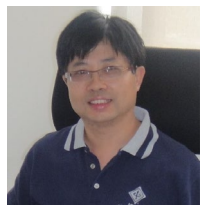

**Changyun Wen (Fellow, IEEE)** received the B.Eng. degree from Xi'an Jiaotong University, China, in 1983 and the Ph.D. degree from the University of Newcastle, Australia in 1990. From August 1989 to August 1991, he was a Postdoctoral Fellow at University of Adelaide, Australia. Since August 1991, he has been with Nanyang Technological University, Singapore, where he is currently a Full Professor. His main research activities are in the areas of control systems and applications, cyber-physical systems, smart grids, complex systems and networks. Some of his publications in these areas are available in https://publons.com/researcher/2816818/changyun-wen/publications/.

Prof Wen is a Fellow of IEEE and Fellow of the Academy of Engineering, Singapore. He was a member of IEEE Fellow Committee from January 2011 to December 2013 and a Distinguished Lecturer of IEEE Control Systems Society from 2010 to 2013. Currently he is the co-Editor-in-Chief of IEEE Transactions on Industrial Electronics, Associate Editor of Automatica (from Feb 2006) and Executive Editor-in-Chief of Journal of Control and Decision. He also served as an Associate Editor of IEEE Transactions on Automatic Control from 2000 to 2002, IEEE Transactions on Industrial Electronics from 2013 to 2020 and IEEE Control Systems Magazine from 2009 to2019. He has been actively involved in organizing international conferences playing the roles of General Chair (including the General Chair of IECON 2020 and IECON 2023), TPC Chair (e.g. the TPC Chair of Chinese Control and Decision Conference since 2008) ect.

He was the recipient of a number of awards, including the Prestigious Engineering Achievement Award from the Institution of Engineers, Singapore in 2005, and the Best Paper Award of IEEE Transactions on Industrial Electronics in 2017.

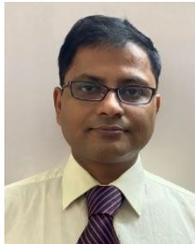

**Suvajit Mukherjee (Senior Member, IEEE)** received the B.E. degree in electrical engineering from Indian Institute of Engineering Science and Technology, Shibpur, India, in 2001, the M.E. degree in electrical engineering from Jadavpur University, Kolkata, India, in 2004 and the Ph.D. degree in electrical engineering from the Indian Institute of Technology, Kharagpur, India, in 2008.

From 2007 onwards he worked in various power industries like Emerson Network Power, Optimal Power Solutions, Stesalit Limited, and is presently working in Rolls-Royce Singapore. His fields of interests are multi-level converters, high voltage converters, control of high-power drives, sensorless ac motor control, UPS, solar inverters and battery charging. He has great experience for electrification architectures for Aerospace, Marine and land-based systems. He is a senior member of IEEE from 2015 and has received numerous accolades and awards from industry.

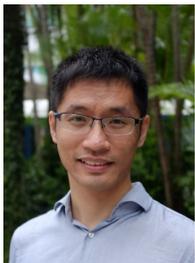

**Xin Zhang (Senior Member, IEEE)** received the Ph.D. degree in Automatic Control and Systems Engineering from the University of Sheffield, U.K., in 2016 and the Ph.D. degree in Electronic and Electrical Engineering from Nanjing University of Aeronautics & Astronautics, China, in 2014.

From February 2017 to December 2020, he was an Assistant Professor of power engineering with the School of Electrical and Electronic Engineering, Nanyang Technological University, Singapore. Currently, he is the Professor at Zhejiang University. He is generally interested in power electronics, power systems, and advanced control theory, together with their applications in various sectors.